\documentclass[12pt]{article}
\usepackage[latin1]{inputenc}
\usepackage{amsfonts}
\usepackage{amssymb,amsmath}

\def\ep{\varepsilon}

\font\ss=cmss10
\def\LiE{{\ss L\kern-.25em\raise 0.59ex\hbox{\i}\kern 0.03em E}}

\def\p{\partial}

\def\da{{\dot a}}
\def\db{{\dot b}}
\def\ep{\varepsilon}
\def\mn{{\mu\nu}}

\def\({\left\{}
\def\){\right\}}
\def\[{\left[}
\def\]{\right]}

\def\Proy{\mathbb{P}}

\newcommand{\cor}[1]{\left\{#1\right\}}
\newcommand{\dir}[1]{\left\{#1\right\}_{\textrm{D}}}
\newcommand{\con}[1]{\left[#1\right]}

\newcommand{\du}[2]{_{#1}^{\phantom{#1}#2}}
\newcommand{\ud}[2]{^{#1}_{\phantom{#1}#2}}
\newcommand{\mat}[1]{\left(\begin{matrix}#1\end{matrix}\right)}
\newcommand{\ket}[1]{\left|#1\right>}

\newcommand{\ffrac}[2]{\genfrac{}{}{0pt}{}{#1}{#2}}

\title{Superprojectors in $D=10$} \vskip 5.5cm
\author{Nicolás Hatcher\\ \vspace{2 cm}\\ A. Restuccia, J. Stephany} 
\date{}

\begin{document}
\maketitle
\thispagestyle{empty}
\vspace{-10cm} \hfill{Preprint {\bf SB/F/06-338}} \hrule
\vspace{5.5cm}
\begin{center}
\textit{ Universidad Simón Bolívar, Departamento de Matemáticas, \\
Apartado Postal 89000, Caracas 1080-A, Venezuela.}\vspace{2cm}\\
\textit{Universidad Simón Bolívar, Departamento de Física\\
Apartado 89000, Caracas 1080A, Venezuela.}\vspace{1.5cm}\\
\textit{nhatcher@fis.usb.ve, arestu@usb.ve, stephany@usb.ve}
\end{center}
\begin{abstract}
We classify all massive irreducible representations of super Poincaré in $D=10$.
New Casimir operators of super Poincaré are presented whose eigenvalues completely specify the representation. It is shown that a scalar superfield contains three irreducible representations of massive supersymmetry and we find the corresponding superprojectors. We apply these new tools to the quantization of the massive superparticle and we show that it must be formulated in terms of a superfield $B_\mn$ satisfying an adequate covariant restriction.
\end{abstract}

\vskip 2cm \hrule
\bigskip
\centerline{\bf UNIVERSIDAD SIMÓN BOLÍVAR} \vfill
\section{Opening words}
In four dimensions a massive scalar superfield can be written in a unique form as a chiral, an antichiral and a transverse superfield. In this work we describe the analog of this in $D=10$. We shall see that a scalar superfield $\Psi(x^\mu,\theta^a)$ can be written in a unique way as the sum of three superfields that correspond to the 44, 84 and 128 irreducible representations of supersymmetry. We explicitly construct superprojectors to this three subspaces. 

In section 2 we put forward the massive supermultiplets in $D=10$ and show that a given representation of $SO(9)$ \textit{induces} a massive representation of super Poincaré. Starting from the Casimir of Zumino and Pasqua \cite{ZumPas} we construct a set of four Casimir operators of super Poincaré whose eigenvalues specify the irreducible representation of super Poincaré.
In section 3 we use this information to discover the massive irreducible representations inside a scalar superfield. Using the previously constructed Casimir operators we find the superprojectors to this irreducible representations and write them in terms of the covariant derivatives.
In section 4 we employ these new tools to solve the non-trivial problem of the covariant quantization of massive superparticles in ten dimensions. 
Section 5 is the mandatory section devoted to draw some conclusions.
\section{Massive supermultiplets in $D=10$}
Before beginning our search for superprojectors we need to know what can be found inside a superfield. That is, we need to classify all the irreducible representations (irreps) of super Poincaré\\
\textbf{Theorem}\\
Massive irreducible representations of the super Poincaré algebra are labelled by an irreducible representation of $SO(9)$, we call $R$ and the mass. The spin content of such representation can be obtained from the product decomposition $R\otimes 44\oplus R\otimes 84\oplus R\otimes 144$\\
\textbf{Proof}\\
Suppose we have a space $\mathcal H$ that supports an irreducible representation of super Poincaré. Select the finite dimensional subspace $\mathcal H_m$ that satisfies the conditions
\begin{gather}
P_\mu \ket{v}=\delta_{\mu,0}m\ket{v}\qquad \forall \ket{v}\in \mathcal V
\end{gather}
This space \textit{naturally} carries a representation of $SO(9)$. The algebra of the supercharges now takes the form
\begin{gather}
\cor{Q_a,\bar Q_\db}=2m\delta_{a\db}
\end{gather}
Note that $\bar Q_\db=J\du{\db}{a}Q_a$ and we cannot impose the condition $Q_a\ket{v}=0$ as in $D=4$. We then define the creation an annihilation operators
\begin{gather}
b_\alpha=\frac{1}{\sqrt{2m}}\left(Q_{\alpha}+iQ_{8+\alpha}\right)\nonumber\\
b^\dag_i=\frac{1}{\sqrt{2m}}\left(\bar Q_\alpha-i\bar Q_{8+\alpha}\right) \\
\alpha=1,...,8\nonumber
\end{gather}
In $\mathcal H_m$ let us consider the set of states $\ket{v}\in\mathcal V$ that satisfy
\begin{gather}
b_\alpha\ket{v}=0
\end{gather}
We call $\mathcal V$ the \textit{characterizing} space. Observe that any state in $\mathcal H_m$ can be written as $(b^\dag_\alpha)^k\ket{v}$ for some $\ket{v}\in V$ with  $0\leq k\leq 8$. In this way $\mathcal H_m$ is decomposed into a tensor product $\mathcal H_m=\mathcal V\otimes\mathcal T$, where $\mathcal T$ is a $2^8$ dimensional vector space build from a vector in $\mathcal V$ and the creation operators.\\
Neither $\mathcal V$ nor $\mathcal T$ are invariant under the transformation generated by $J_{ij}$.  But we shall see that there is a representation of $SO(9)$ acting in $\mathcal V$, with generators $V_{ij}$ and another representation acting on $\mathcal T$ with generators $T_{ij}$ such that the representation defined by $J_{ij}$ can be considered the tensor product of this two representations. To prove this we construct the following operator in the whole space  \cite{ZumPas}
\begin{gather}
C_{\mu\nu\rho}=J_{[\mn}P_{\rho]}+\frac{1}{24i}QS_{\mn\rho}Q\\
C_\mn=P^\rho C_{\mu\nu\rho}
\end{gather}
with the important property $\con{Q_a,C_\mn}=0$. In other words the operator $C_\mn$ fulfils the condition $C_\mn\mathcal V\in\mathcal V$. As far as the $\mathcal V$ subspace is concerned the operator is $C_{ij}$. We define, following \cite{ZumPas}, $V_{ij}=-\frac{3}{m^2}C_{ij}$. The algebra of this operators is the algebra of  $SO(9)$. The other generators are $T_{ij}=\frac{-1}{6im}QS_{ij0}Q$. It is easy to see that $\con{T_{ij},V_{lk}}=0$ because $\con{Q_a,C_\mn}=0$ so they act on different spaces. The next step is to prove that the representation of $SO(9)$ provided by $T_{ij}$ is the direct sum $44\oplus 84\oplus 128$. There is a part of the $J_{ij}$ that does not mix the creation and annihilation operators, this is the $SU(4)\times SU(2)$ subgroup of $SO(9)$ (see \cite{Slan}). So we may say that the index $\alpha$ transforms in the $(4,2)$ representation of $SU(4)\times SU(2)$ or, what is the same in the $[0,0,1,1]$ representation (the first three numbers indicate the Dynkin label for the $SU(4)=A_3$ representation and the last one is the Dynkin label for $A_1$). To see the irreducible representations of $SU(4)\times SU(2)$ in $\mathcal T$ we need to know what irreducible representation appear in the antisymmetric product of $[0,0,1,1]^k$ for $k=0,..,8$. This is done quite easily with the program \LiE\cite{LiE}.
\begin{align*}
(\[0,0,1,1\])^0_A &  \longrightarrow \[0,0,0,0\]\\
\[0,0,1,1\] & \longrightarrow \[0,0,1,1\]\\
(\[0,0,1,1\]\times \[0,0,1,1\])_A  & \longrightarrow   \[0,0,2,0\]+\[0,1,0,2\]\\
(\[0,0,1,1\])^3_A  & \longrightarrow  \[0,1,1,1\]+\[1,0,0,3\]\\
(\[0,0,1,1\])^4_A  & \longrightarrow  \[0,0,0,4\]+\[0,2,0,0\]+\[1,0,1,2\]\\
(\[0,0,1,1\])^5_A  & \longrightarrow  \[0,0,1,3\] +\[1,1,0,1\]\\
(\[0,0,1,1\])^6_A  & \longrightarrow  \[0,1,0,2\]+\[2,0,0,0\]\\
(\[0,0,1,1\])^7_A  & \longrightarrow   \[1,0,0,1\]\\
(\[0,0,1,1\])^8_A  & \longrightarrow  \[0,0,0,0\]
\end{align*}
Studying the branching rules of $SO(9)\rightarrow SU(4)\times SU(2)$ irreps \cite{Slan} we are able to rearrange them into irreps of $SO(9)$
\begin{align*}
\ffrac{[0,0,0,2]}{44}& \ffrac{=}{ }\ffrac{[0,0,0,0]}{(1,1)}\ffrac{+}{ }\ffrac{[0,0,0,4]}{(1,5)} \ffrac{+}{ }\ffrac{[0,1,0,2]}{(6,3)} \ffrac{+}{ } \ffrac{[2,0,0,0]}{(20,1)}\\
\ffrac{[0,0,1,0]}{84}& \ffrac{=}{ }\ffrac{[0,0,0,0]}{(1,1)}\ffrac{+}{}\ffrac{[0,2,0,0]}{(10,1)}\ffrac{+}{ } \ffrac{[0,0,2,0]}{(10,1)} \ffrac{+}{ }\ffrac{[0,1,0,2]}{(6,3)} \ffrac{+}{ }\ffrac{[1,0,1,2]}{(15,3)}\\
\ffrac{[1,0,0,1]}{128}& \ffrac{=}{ }\ffrac{[0,0,1,1]}{(4,2)}\ffrac{+}{}\ffrac{[1,0,0,1]}{(4,2)}\ffrac{+}{ } \ffrac{[1,0,0,3]}{(4,4)} \ffrac{+}{ }\ffrac{[0,0,1,3]}{(4,4)} \ffrac{+}{ }\ffrac{[1,1,0,1]}{(20,2)}\ffrac{+}{ }\ffrac{[0,1,1,3]}{(20,2)}
\end{align*}
where we have written each branching in two different notations. The first line shows how the Dynkin labels of $SO(9)$ break down in Dynkin labels of $SU(4)\times SU(2)$. The second line in each case merely indicates the dimension.\\
So, the characterizing space carries a representation of $SO(9)$, call it $R$, and the creation operators fit into irreducible representations of $SO(9)$, $(44,84,128)$. The complete set of irreducible representation of $SO(9)$ inside a super Poincaré representation (the spin content) is given by the decomposition of the products $(44\otimes R,84\otimes R,128\otimes R)$.\\
For example if we start with the $[0,0,0,0]$ irrep of $SO(9)$ we obtain the smallest irrep of super Poincaré, whose field content is a 44 irrep of Poincaré, a 84 irrep of Poincaré, and the 128 irrep of Poincaré. We will call this the 1 irrep of super Poincaré. Note that although we start from an irrep of Poincaré to build and irrep of super Poincaré just as in $D=4$ in this case there is no relation between the irrep we start with and the spin content of the irrep we obtain.\\
The following table shows the spin content of some massive supermultiplets\\

\begin{table}[!ht]
\begin{center}
\begin{tabular}{|c|p{6 cm}|p{6 cm}|}
\hline
Irrep & Bosonic fields & Fermionic Fields  \\
\hline
1     & 44+84 & 128    \\
\hline
9 & 9+36+126+156+231+594 & 16+128+432+576\\
\hline
16 & $9+36+44+84+126+231+594+924$ & $16+2\times 128+432+576+768$     \\
\hline
36 & $9+36+44+84+126+231+594+910+924+1650$ & $16+2\times 128+432+576+768+2560$    \\
\hline
44 & $1+36+44+84+231+45+495+924+2457$ & $16+128+432+576+1920+2560$     \\
\hline
84 & $1+36+44+2\times 84+126+231+495+594+2\times 924+1980+2457+2772$ & $16+2\times 128+2\times 432+576+672+768+2560+5040$  \\
\hline
128 & $1+9+2\times 36+44+2\times 84+2\times 126+156+2\times 231+495+2\times 594+910+2\times 924+1650+2457+2772+3900$ & $2\times 16+3\times 128+3\times 432+2\times 576+672+ 768+1920+2\times 2560+5040$   \\
\hline
\end{tabular}
\caption{Massive supermultiplets in $D=10$}\label{tabla:supermultiplets}
\end{center}
\end{table}
To close this section we would like to make the following observation regarding the Casimir invariants of the super Poincaré algebra. There are four Casimir operators of $SO(9)$ and since every representation of supersymmetry can be build from a representation of $SO(9)$ it is natural to expect that there would be some equivalent invariants. In fact they are constructed from $C_\mn$ as follows
\begin{gather}
C_2=C_\mn C^{\nu\mu}\qquad C_4=C_{\mu_1\mu_2}C^{\mu_2\mu_3}C_{\mu_3\mu_4}C^{\mu_4\mu_1}\\
C_6=C_{\mu_1\mu_2}\cdots C_{\mu_{11}\mu_{12}}C^{\mu_{12}\mu_1}\qquad C_8=C_{\mu_1\mu_2}\cdots C^{\mu_{16}\mu_1}
\end{gather}
To calculate their eigenvalues the safest way is to compute them in the characterizing space $\mathcal V$. There $C_{ij}=-\frac{m^2}{3}V_{ij}$ and so the eigenvalues of $C_{2p}$ are $\frac{m^{4p}}{3^{2p}}I_{2p}$.
\section{Superprojectors}
Once we have classified all the irreducible representations of supersymmetry our next commitment is to understand what irreducible representations of supersymmetry dwell inside a scalar superfield. This problem is more or less easy to solve if we rely on our previous work. The solution is given by the following general principle, valid in all dimensions, any number of supercharges, and for a superfield with any kind of index:\\
\textit{The field content inside a superfield is the same that the spin content inside a representation of supersymmetry.} \\
Let us explain a little more the content of this useful statement. In four dimensions and $N=1$, the smallest supersymmetry representation contains two scalars and a spin $1/2$ particle, in turn the scalar superfield contains two superspin zero representations and one superspin $1/2$.\\
Let us transliterate this to $D=10$. In ten dimensions the smallest supersymmetry representation has a 44, an 84 and a 128 irreducible representation of Poincaré and a scalar superfield has a 44, an 84 and a 128 irreducible representations of super Poincaré\\
The reason for this is, of course, that the algebra of the covariant derivatives \textit{is the same} that the algebra of the supercharges and we can proceed exactly as before, splitting the derivatives and then rearranging them into $SO(9)$ representations.\\
In four dimensions there are two competitive mechanisms to construct superprojectors. One takes advantage of chirality \cite{GatSie} and the other tries to find a complete set of Casimir operators  \cite{Sok} \cite{RitSok}. The first method turns out to be more successful because in a general superfield a given representation occurs more than once and for $N=3,4$ there are simply not enough invariants. It is curious that in $D=10$ it is the second method that is succesful because there is no chirality condition.\\
All that we need is a Casimir of super Poincaré $K$ that has three different eigenvalues $\lambda_1$, $\lambda_2$ and $\lambda_3$ for the 44, 84 and 128 representations, then the three projectors are given by:
\begin{gather}
\Proy_1=\frac{1}{(\lambda_1-\lambda_2)(\lambda_1-\lambda_3)}(K-\lambda_2 I)(K-\lambda_3 I)\\
\Proy_2=\frac{1}{(\lambda_2-\lambda_1)(\lambda_2-\lambda_3)}(K-\lambda_1 I)(K-\lambda_3 I)\\
\Proy_3=\frac{1}{(\lambda_3-\lambda_1)(\lambda_3-\lambda_2)}(K-\lambda_2 I)(K-\lambda_1 I)
\end{gather}
A simple look a table \ref{tabla:SO9cas} shows that the $C_2$ Casimir is not good enough but the $C_4$ Casimir satisfy our needs.\\
In superspace these projectors can be written, as in $D=4$ solely in terms of covariant derivatives since,
\begin{gather}
C_\mn=P^\rho \left(J_{[\mn}P_{\rho]}+\frac{1}{24i}QS_{\mn\rho}Q\right)=-\frac{1}{24i}P^\rho DS_{\mu\nu\rho}D
\end{gather}
Note that $L_{[\mn}P_{\rho]}=0$, where $L_\mn=X_\mu P_\nu-X_\nu P_\mu$, then use (\ref{Qa}) and (\ref{Da}) to write $\Pi_a$ and $\Theta^a$ in terms of $Q_a$ and $D_a$.\\
A further simplification of the superprojectors can be achieved by noting that $W_2$ can be used to build a superprojector into the 84 irreducible representation, so that $\Proy_{84}$ is simply written as
\begin{gather}
\Proy_{84}=\frac{C_2-\alpha}{\beta-\alpha}=\frac{9C_2-28P^4}{8P^4}\, ,
\end{gather}
where $\alpha=\frac{28P^4}{9}$ is the eigenvalue for the 44 or the 128 representations and $\beta=4P^4$ is the eigenvalue for the $84$ representation. Of course the $C_4$ can also be used to construct $\Proy_{84}$. This gives us a relation between $(W_4)^2$, $W_4$ and $W_2$ which shows that the three superprojectors are linear in $W_2$ and $W_4$.\\
Finally let us remark that to find the irreducible representation of a superfield with some index all we have to do is to read the table \ref{tabla:supermultiplets}. For example the supermultiplets inside a divergenceless, vector superfield $A_\mu$ is $(9\otimes44)\oplus(9\otimes 84)\oplus (9\otimes 128)$, to find the corresponding superprojectors is another game. For a concrete example see the next section.
\section{Application: Quantization of a superparticle}
The initial motivation for this study came from the observation in \cite{SteResHat} that with a superprojector in ten dimensions we might have a chance to describe the correct quantum mechanics of massive superparticles. In this section we apply the new tools we have found to this problem. We shall see that the knowledge of the superprojectors provides us with a powerful mechanism to select irreducible representations inside a superfield.\\
We start from an action for a massive superparticle in ten dimensions
\begin{gather}
S=\frac{1}{2}\int (e^{-1}\omega^\mu\omega^\nu\eta_\mn-m^2 e)d\tau
\end{gather}
with $\omega^\mu=\dot x^\mu+i\theta^a S\ud{\mu}{ab}\dot \theta^b$. In this system we have a first class constraint related to the invariance under reparametrization and a family of second class constraints.
\begin{gather}
d_a=\pi_a+ip_\mu S\ud{\mu}{ab}\theta^b\\
\cor{d_a,d_b}=-2ip_\mu S\ud{\mu}{ab}
\end{gather} 
Because of the presence of second class constraints the standard algorithm instructs us to substitute the Poisson brackets with Dirac brackets. As a result we obtain the algebra
\begin{gather}
\dir{\theta^a,\theta^b}=\frac{i}{2p^2}p_\mu S^{\mu\,ab}\\
\dir{\theta^a,x^\mu}=\frac{1}{2p^2}\theta^b S\ud{\mu}{bc}S^{\nu\,ca}p_\nu\\
\dir{x^\mu,x^\nu}=\frac{-\Sigma^\mn}{p^2}
\end{gather}
where
\begin{gather}
J^\mn=L^\mn+\Sigma^\mn\\
L^\mn=x^\mu p^\nu-x^\nu p^\mu\qquad \Sigma^\mn=\frac{-1}{4}\theta S^\mn\pi\\
\pi_a=-ip_\mu S\ud{\mu}{ab}\theta^b
\end{gather}
Straightforward canonical quantization now demands to switch Dirac brackets by commutators. So that our problem is now to find a set of operators that fulfil the quantum algebra
\begin{gather}
\cor{\hat\Theta^a,\hat\Theta^b}=\frac{-1}{2P^2}P_\mu S^{\mu\,ab}\\
\con{\hat X^\mu,\hat\Theta^a}=\frac{i\hat \Theta^b}{2P^2}P_\nu S\ud{\mu}{bc}S^{\nu\, ca}\\
\con{\hat X^\mu,\hat X^\nu}=\frac{-i\Sigma^\mn}{P^2}
\end{gather}
where $\Sigma^\mn$ is the \textit{internal} angular momentum given in this case by
\begin{gather}
\Sigma_\mn=\frac{-1}{4}\hat\Theta S_\mn\hat\Pi\label{eq:internal}
\end{gather}
It is not difficult to show that for this algebra the invariant $C_2$ defined above is zero, indeed $C_\mn$ is zero because it is zero in te characterizing space so it is zero everywhere by supersymmetry. That is, a quantum superparticle is described by a 1 irrep of super Poincaré. But we would like to implement the algebra as a set of operators acting in Hilbert space of superfields.\\
The idea in \cite{SteResHat2} was that this algebra can be implemented at the quantum level if we find a superprojection operator $\Proy$ that meet the requirements
\begin{gather}
\con{\Proy, Q_a}=\con{\Proy,P_\mu}=\con{\Proy,J_\mn}=0\\
\Proy D_a\Proy=0
\end{gather}
Then a set of operators $(\hat X^\mu,\hat \Theta^a)$ that satisfy the quantum algebra of superspace would be given by the rule
\begin{gather}
\hat X^\mu= \Proy X^\mu\Proy\qquad \hat\Theta^a= \Proy\Theta^a\Proy
\end{gather}
We now have three such projectors and hence we have three representations of the algebra. The problem is that for these representations the internal angular momentum is \textit{not} given by (\ref{eq:internal}) but has a more complex expression \cite{SteResHat2}
\begin{gather}
\Sigma_\mn=\frac{-1}{4}\hat\Theta S_\mn\hat\Pi+T_\mn\\
T_\mn=\frac{-1}{4}\Proy \Theta S_\mn\Pi\Proy+\frac{1}{4}\hat\Theta S_\mn\hat\Pi=\frac{P_\alpha}{32P^2}\Proy D S^\alpha S^\mn D\Proy
\end{gather}
This extra term can also be written in terms of the operator defined earlier \cite{SteResHat2} $C_\mn$ as $T_\mn=P^\alpha P_{[\alpha} C_{\mn]}$.
In four dimensions there are projectors (for a scalar superfield) such that $T_\mn=0$, but this is no longer true in $D=10$. For the three projectors we have found the term $C_\mn$ is strictly non-zero. If we want to realize the algebra we need to consider a \textit{different} superfield. Now table \ref{tabla:supermultiplets} comes handy. To get the correct algebra we need the smallest supermultiplet. The simplest superfield which contains such a representation is a symmetric, traceless and divergenceless tensor $B_\mn(x^\mu,\theta^a)$ that transforms in the 54 irrep of $SO(9,1)$. The 54 irrep decomposes as the 44+9+1 irreps of $SO(9)$ so that the irreps of super Poincaré inside this superfield are given by the $(44\oplus 9\oplus 1)\otimes(44\oplus 84\oplus 128)$. The 1 representations occurs only once (in the decomposition of $44\otimes 44$). For this superfields the generators of super Poincaré take the same form that for a scalar superfield except the generators of Lorentz transformations:
\begin{gather}
J_\mn=L_\mn+\Sigma_\mn+R_\mn
\end{gather}
The new term $R_\mn$ is responsible for the transformation of the indices of the superfield. As before we can construct the tensor
\begin{gather}
C_\mn=P^\rho \left(J_{[\mn}P_{\rho]}+\frac{1}{24i}QS_{\mn\rho}Q\right)=P^\alpha\left(R_{[\mn}P_{\alpha]}-\frac{1}{24i}DS_{\mn\alpha}D\right)
\end{gather}
And the corresponding Casimir $C_2$. We know that inside a $B_\mn$ superfield there is a 1 irrep of super Poincaré (and only one), so we are allowed to set the restriction
\begin{gather}
C_2 B_\mn=0
\end{gather}
We can also construct a projector to this representation. First we need to calculate all the irreps inside $B_\mn$ as we have indicated. Then we compute the eigenvalues of $C_2$, $\lambda_i$. Different irreps inside $B_\mn$ may have the same $\lambda_i$, but that does not disturb us. The projector is then
\begin{gather}
\Proy=\frac{\prod_i (C_2-\lambda_i)}{\prod_i(-\lambda_i)}
\end{gather}
Every projector $\Proy$ satisfies $\Proy D_a\Proy=0$ and that is all we need to understand that we have a correct covariant quantization of the massive super particle in $D=10$. Indeed, for the same reason as before, not only $C_2$ is zero but also $C_\mn$. To avoid confusion let us rename the indices of the superfield $B_\mn$ by $B_M$ so that the Lorentz generators may be written as $(R_\mn)\du{M}{N}$. The equations for a superfield describing first quantized superparticles are then
\begin{gather}
\frac{-1}{24i}DS_{\mn\rho}D P^\rho B_M(x,\theta)=P^\rho (P_{[\rho}R_{\mn]})\du{M}{N} B_N(x,\theta)\label{eq:proc1}\\
(P^2+m^2)B_M(x,\theta)=0\label{eq:proc2}
\end{gather}
A rapid look at these equations reveals its non-trivial character. In four dimensions a massive spin 1 particle is described by Proca equations
\begin{gather}
(P^2+m^2)A_\mu=0\qquad \p^\mu A_\mu=0
\end{gather}
If $m\neq 0$ they are equivalent to $\p^\mu F_\mn=m^2 A_\nu$ and the limit $m=0$ of this equation gives Maxwell equations. Something similar happens with supersymmetry. Super Proca equations are
\begin{gather}
D^2 V=\bar D^2 V=0\qquad (P^2+m^2)V=0
\end{gather}
For massive particles this equations are equivalent to
\begin{gather}
\left[\frac{1}{16}\left(\ep^{ab}D_a\bar D^2 D_b+\ep^{\da\db}\bar D_\da D^2\bar D_\db\right)+m^2\right]V=0
\end{gather}
In the limit $m=0$ this reproduces the equations of super Maxwell.
In view of this it is tempting to think that there exists a limit of (\ref{eq:proc1}-\ref{eq:proc2}) which reproduces super Maxwell.
\section{Concluding words}
In 1977 ten dimensional super Yang-Mills theories were discovered. Since then, an immense amount of work has been devoted to its study and comprehension. Many secrets of those theories have been uncovered but there seems to be many more yet to be discovered. To date not entirely satisfactory explanation exits for the off-shell structure of super Yang-Mills theories. Whatever the solution of the problem will turn out to be we are confident that the present work represents a necessary step in that direction.\\
We have classified all massive multiplets in ten dimensions (see table \ref{tabla:supermultiplets}). With this information we have found the supermultiplets inside a scalar superfield. More important, we have constructed superprojectors to this supermultiplets $\Proy_{44}$, $\Proy_{84}$ and $\Proy_{128}$ in terms of covariant derivatives. In other words we have found the possible restrictions that can be imposed covariantly on a scalar superfield. Our equations are the closest one can get in $D=10$ to a chirality condition whose importance in $D=4$ is difficult to overestimate. As an example we have found the correct quantization of massive superparticles in ten dimensions providing a covariant equation for a superfield $B_\mn$ whose solution is the 1 supermultiplet. The zero mass limit of equations (\ref{eq:proc1}) and (\ref{eq:proc2}) should be intimately related to the problem of the covariant quantization of the massless superparticle \textit{and} the off-shell structure of super Yang-Mills theories. But the precise relationship, if it exist, is deferred to future work.
\section{Acknowledgments} 
This work was supported by Did-Usb grants Gid-30 and Gid-11 and by Fonacit grant G-2001000712.  For N.H. it is a pleasure to thank S. J. Gates, Jr. for a discussion and encouragement.

\setcounter{equation}{0}
\renewcommand{\theequation}{\thesection.\arabic{equation}}
\appendix
\section{Gamma matrices}
Here we fix conventions and notations. The metric is $\eta_\mn=\textrm{diag}(-1,+1,...,+1)$. Gamma matrices satisfy
\begin{gather}
\cor{\Gamma^\mu,\Gamma^\nu}=-2\eta^\mn
\end{gather}
the indices run $\mu=0,..,9$. We find a representation in which this matrices are antidiagonal
\begin{gather}
\Gamma^\mu=\mat{0 & S^\mu\\ \bar S^\mu & 0}
\end{gather}
with $S^0=\bar S^0=I$ and $S^i=-\bar S^i$. The matrix are constructed inductively if $\Gamma^\mu$ are the matrices in $D$ dimensions $\tilde \Gamma^\mu$ are Dirac matrices in $D-2$ we have
\begin{gather}
S^i=\tilde\Gamma^i\tilde W\qquad i=1,2,...,D-3\\
S^{D-2}=i\tilde W\tilde\Gamma^0,\quad S^{D-1}=\tilde{W}\\
\tilde W=\mat{I & 0\\ 0 & -I}
\end{gather}
where $\tilde W$ is the Weyl matrix in $D-2$. A compatible selection of indices is $S\ud{\mu}{a\db}$ and $\bar S^{\db a}$, where a dot means they transform in the conjugate representation, $a=1,...,16$. In $D=10$ there is a matrix $J\du{a}{\db}$ that converts one index into another. We define the symmetric matrices $S\ud{\mu}{ab}=S\ud{\mu}{a\db}J\du{b}{\db}$ and hope there is no confusion. All spinors are Majorana $\lambda^a=J\ud{a}{\db}\bar \lambda^\db$. We define
\begin{gather}
{S_\mn}\du{a}{b}=S_{\mu\,a\db}\bar S\du{\nu}{\db b}-S_{\nu\,a\db}\bar S\du{\mu}{\db b}\\
S_{\mn\rho\,ab}=\frac{1}{3!}\left(S_{\mu\,a\db}\bar S\du{\nu}{\db c}S_{\mu\,c\dot c}J\du{b}{\dot c}\pm \textrm{antisym}\right)
\end{gather}
Note that $S_{\mn\rho}$ is antisymmetric in $a,b$. The generators of Poincaré are $\frac{i}{4}S_\mn$.\\
Superspace is parameterized by $(x^\mu,\theta^a)$. In this superspace the representation of supersymmetry  and covariant derivatives is
\begin{gather}
P_\mu=-i\p_\mu\qquad \Pi_a=-i\p_a
J_\mn=X_\mu P_\nu-X_\nu P_\mu-\frac{1}{4}\Theta S_\mn \Pi\\
Q_a=\Pi_a-iP_\mu S\ud{\mu}{ab}\Theta^b\label{Qa}\\
D_a=i\Pi_a-P_\mu S\ud{\mu}{ab}\Theta^b\label{Da}
\end{gather}
Important relations
\begin{gather}
\con{J_\mn,Q_a}=-\frac{i}{4}{S^\mn}\du{a}{b}Q_b\\
\cor{Q_a,Q_b}=\cor{D_a,D_b}=2P_\mu S\ud{\mu}{ab}
\end{gather}
\section{Some group theory}
$SO(9)$ is the Lie group $B_4$. An irreducible representation is given by four numbers $[a_1,a_2,a_3,a_4]$, the Dynkin labels. The dimension of this representation is computed by Weyl's formula (see \cite{Oku}). A representation will be some times named after the dimension or after the Dynkin labels. Racah \cite{Rac} has shown that there are four independent Casimir operators for this group that we can write as
\begin{gather}
I_2=J_{ij}J^{ji}\qquad I_4=J_{i_1i_2}J^{i_2i_3}J_{i_3i_4}J^{i_4i_1}\\
I_6=J_{i_1i_2}\cdots J^{i_{12}i_1}\qquad I_8=J_{i_1i_2}\cdots J^{i_{16}i_1}
\end{gather}
We will only need the $I_2$ and $I_4$ invariants but we list all for the readers convenience. The value of the invariant $I_{2p}$ was computed in a beautiful paper by S. Okubo \cite{Oku} and the result is (for us $n=4$, these formula are valid for $SO(2n+1)$)
\begin{gather}
I_{2p}=\sum_{j=1}^n\left\{\left(m_j+\frac{1}{2}-n\right)^{2p}\frac{m_j+1}{m_j}{\prod_{\ffrac{k=1}{k\neq j}}^n}  \frac{(m_j+1)^2-(m_k)^2}{(m_j)^2-(m_k)^2}+\right.\nonumber\\ \left.
\left(m_j-\frac{1}{2}+n\right)^{2p}\frac{m_j-1}{m_j} {\prod_{\ffrac{k=1}{k\neq j}}^n}  \frac{(m_j-1)^2-(m_k)^2}{(m_j)^2-(m_k)^2}\right\}+n^{2p}
\end{gather} 
where 
\begin{gather}
l_k=\frac{a_n}{2}+\sum_{i=k}^{n-1} a_i\qquad k\neq n\nonumber\\
l_n=\frac{a_n}{2}\nonumber\\
m_i=l_i+n-i+\frac{1}{2}
\end{gather}
With this data we can fill the table \ref{tabla:SO9cas}.

\begin{table}[!ht]
\begin{center}
\begin{tabular}{|c|c|c|c|c|c|}
\hline
Dynkin & Dimension & $I_2$ & $I_4$ & $I_6$ & $I_8$ \\
\hline
$[1,0,0,0]$     & 9 & 16    & 464   & 29136     &1864144  \\
\hline
$[0,0,0,1]$ & 16 & 18     & 256.5  & 4096.125     & 65536.03125 \\
\hline
$[0,1,0,0]$ & 36 & 28     & 644  &   29568   & 1441804 \\
\hline
$[2,0,0,0]$ & 44 & 36     & 1404  & 108936   & 8805924\\
\hline
$[0,0,1,0]$ & 84 & 28     & 644   &   29568   & 1441804\\
\hline
$[0,0,0,2]$ & 126 & 40     & 680  & 14520     & 325960\\
\hline
$[1,0,0,1]$ & 128 & 36     & 931.5   & 51291     & 3471782.90625 \\
\hline
\end{tabular}
\caption{Irreducible representation of $SO(9)$}\label{tabla:SO9cas}
\end{center}
\end{table}
Some values are easy to calculate, for example the value of the $I_{2p}$ Casimir for the spinor representation can be computed form Clifford algebra and the result is \\ \mbox{$\frac{D-1}{2^{2p}}\left((D-1)^{2p-1}+1\right)$} that in the special case of $D=9$ reduces to $2^{4p}+\frac{1}{2^{2p-3}}$. The value of a Casimir for a given representation $R$ is denoted by $I_{2p}^R$. For example $I_6^{126}=14520$.\\
Irreducible representations of Poincaré $\mathcal P=SO(9,1)\otimes_S T_4$ can be obtained from those of $SO(9)$ as usual. From the $I_{2p}$ invariants we build a Casimir for $\mathcal P$. First define the analog of the Pauli-Lubanski four vector
\begin{gather}
W_{\mn\rho}=J_{[\mn}P_{\rho]}=\frac{2}{3!}\left(J_{\mn}P_\rho+J_{\nu\rho}P_\mu+J_{\rho\mu}P_\nu\right)\\
W_\mn=W_{\mn\rho}P^\rho
\end{gather}
This tensor commutes with $P_\mu$ so we can build Casimir invariants in complete analogy with the $I_{2p}$ operators
\begin{gather}
W_2=W_{\mn}W^{\nu\mu}\qquad W_4=W_{\mu_1\mu_2}W^{\mu_2\mu_3}W_{\mu_3\mu_4} W^{\mu_4\mu_1}\nonumber\\
W_6=W_{\mu_1\mu_2}\cdots W^{\mu_{12}\mu_1}\qquad W_8=W_{\mu_1\mu_2}\cdots W^{\mu_{16}\mu_1}
\end{gather}
To compute the eigenvalues consider a state $\ket{\Psi}$ that satisfies:
\begin{gather}
P_0\ket{\Psi}=m\ket{\Psi}\qquad P_i\ket{\Psi}=0\qquad i=1..9
\end{gather}
For this state we have
\begin{gather}
\left.
\begin{matrix}
W_{ij 0}=\frac{1}{3}m^2J_{ij}\\
W_{ijk}=0\qquad i,j,k=1...9
\end{matrix}\right\}\Rightarrow W_{2p}=-\left(\frac{m^2}{3}\right)^{2p}I_{2p}
\end{gather}
Finally the computer program \LiE \cite{LiE} has been useful in more than one way. For example to compute the antisymmetric product decomposition $([0,0,1,1])^4$ of the group $A_3\times A_1$ all we need to write is
\begin{verbatim}
> alt_tensor(4,[0,0,1,1],A3*A1)
\end{verbatim}
and the machine gives the answer.

\end{document}